\documentclass [12 pt]{article}
\usepackage{psfig}
\usepackage{epsfig}
\newcommand {\be} {\beta}
\newcommand {\al} {\alpha}

\newcommand {\epb} {\bar{\epsilon}}
\newcommand {\Wb} {\bar{W}}
\newcommand {\Delb} {\bar{\Delta}}
\newcommand {\pb} {\bar{p}}
\newcommand {\Ab} {\bar{A}}
\newcommand {\spc} {\;\;\;}
\newcommand {\la} {\lambda}
\newcommand {\De} {\Delta}
\newcommand {\lbr} {\left [}
\newcommand {\rbr} {\right ]}
\newcommand {\lpar} {\left (}
\newcommand {\rpar} {\right )}
\newcommand {\tti} {\tilde{t}}
\newcommand {\uu} {$\_$}

\newcommand {\muma} {\frac{\mu}{m_a}}
\newcommand {\mum} {\frac{\mu}{m}}
\newcommand {\pv} {\vec{p}}
\newcommand {\qv} {\vec{q}}
\newcommand {\Wv} {\vec{W}}
\newcommand {\Dev} {\vec{\Delta}}
\newcommand {\qb} {\bar{q}}

\newcommand {\etb} {\bar{\eta}}
\newcommand {\petb} {\bar{p_{\eta} }}
\newcommand {\Wh} {\hat{W}}
\newcommand {\xh} {\hat{x}}
\newcommand {\xb} {\bar{x}}
\begin{document}
\title{Theory of electron cooling  using electron cooling as an intrabeam
scattering process}
\author{George Parzen}
\date{\today  }
\maketitle
\begin{abstract}
Electron cooling that results when a bunch of electrons overlaps a bunch of ions , 
with both bunches moving at the same velocity,  may be considered to be an 
intrabeam scattering process. The process is similar to 
the usual  intrabeam scattering, 
where the ions scatter from each other and usually results 
in beam growth.  An important 
difference is that in electron cooling the mass of the ion is different from and 
much larger than the mass of the electron. This difference 
considerably complicates  
the intrabeam scattering theory. It introduces a new term in the emittance growth 
rate, which vanishes when the particles are identical and 
their masses are equal, and 
can give rise to emittance cooling of the heavier particles . The term that gives 
rise to beam growth for the usual  intrabeam scattering is also present 
but is much smaller than the cooling term  when  one particle is 
much heavier than the other.
This paper derives the results found for the emittance cooling rates 
due to the scattering of the ions in the ion bunch by the electons 
in the electron bunch. 
\end{abstract}

\section{Introduction}

Electron cooling that results when a bunch of electrons overlaps a bunch of ions , 
with both bunches moving at the same velocity,  may be considered to be an 
intrabeam scattering process. The process is similar to 
the usual  intrabeam scattering, Ref.[1]
where the ions scatter from each other and usually results 
in beam growth.  An important 
difference is that in electron cooling the mass of the ion is different from and 
much larger than the mass of the electron. This difference 
considerably complicates  
the intrabeam scattering theory. It introduces a new term in the emittance growth 
rate, which vanishes when the particles are identical and 
their masses are equal, and 
can give rise to emittance cooling of the heavier particles . The term that gives 
rise to beam growth for the usual  intrabeam scattering is also present 
but is much smaller than the cooling term  when  one particle is 
much heavier than the other.

This paper derives the results found for the emittance cooling rates 
due to the scattering of the ions in the ion bunch by the electons 
in the electron bunch. The derivations given below makes 
considerable use of the results found in two 
previous papers, Ref.[2] and Ref.[3]

\section{The $f(x,p)$ distribution and the scattering rate $\delta N$}

The ions are contained within a bunch 
and their distibution is given by $f_a(x_a,p_a)$ where $N_a f_a(x_a,p_a)$ 
is the number 
of ions in $d^3x_ad^3p_a$. $N_a$ is the number of ions in the bunch.
\[ \int d^3x_ad^3p_a \; f_a(x_a,p_a)=1  \]
The distribution of the electrons in the electon bunch is 
given by $f_b(x_b,p_b)$ and $N_b$ is the number of 
electrons in the electron bunch.
Let $\delta N_a$ be the number of  ions with momentum,  $p_a$ in $d^3p_a$ 
and  space coordinate $x$ in $d^3x$
which are scattered by the electrons  with momentum $p_b$ in $d^3p_b$ 
which are also in  $d^3x$, in the time interval $dt$ , into the 
solid angle $d\Omega'$ corresponding to the
direction $\hat{p_a'}$. Then $\delta N_a$ is given by, Ref.[2],
\begin{eqnarray} 
\delta N_a &=& N_a N_b \sigma_{ab} d\Omega' \frac {d^3p_a}{\gamma_a}
      \frac {d^3p_b}{\gamma_b} f_a(x,p_a)f_b(x,p_b) F(p_a,p_b) d^3x dt \nonumber\\
F(p_a,p_b) &=&  \frac {[(p_ap_b)^2-m_a^2 m_b^2 ]^{1/2}}{m_a m_b}
\end{eqnarray}
$\sigma_{ab}$ is the scattering cross section for the scattering 
of the ions from the electrons.In the expression for $F(p_a,p_b)$, we have put
$c=1$. $F(p_a,p_b)$ has the dimensions of a velocity.

For completeness sake this result is given in the form which is valid in any CS.
For the electron cooling problem for RHIC, one can do all the calclations 
in the Rest CS, which is the CS moving along with the two bunches. In the 
Rest CS, the central particle in either bunch is at rest and the motion of the
motion of the particles may be treated non-reletavistically.In the 
Rest CS , one may put $\gamma_a=\gamma_b=1$ and
   \[ F(p_a,p_b)=|\vec{v_a}-\vec{v_b}|  \]

\section{Growth rates for $<p_{ia}p_{ja}>$}

Following Bjorken and Mtingwa, Ref.[4],  cooling rates will first be  
given for $<p_{ia}p_{ja}>$. where the $<>$ indicate an 
average over all the particles in the bunch. From these one can compute the 
growth rates for the average emittances of the ions, $<\epsilon_{ia}>$.
In a scattering event, where an ion with  momentum $p_a$ scatters off an  
electron with momentum $p_b$, the momenta will change to $p_a'$ and $p_b'$.
Let $\delta p_{ia}$ represent the change in $p_{ia}$ in the collision, 
and similarly for $\delta (p_{ia}p_{ja})$. Then
\begin{eqnarray} 
\delta p_{ia} &=& p_{ia}'-p_{ia}     \nonumber\\
\delta (p_{ia}p_{ja}) &=& p_{ia}' p_{ja}'-p_{ia}p_{ja}
\end{eqnarray}

Using the scattering rate given by Eq.(1), one can now compute 
$ \delta <p_{ia}p_{ja}>$ in the Rest CS,
\begin{eqnarray} 
\delta <(p_{ia}p_{ja}) > &=& N_b \int \spc d^3x d^3p_{a}
      d^3p_{b} f_a(x,p_a) f_b(x,p_b) |\vec{v_a}-\vec{v_b}|   \nonumber\\
      & & \spc  \sigma_{ab} d\Omega' \spc \delta (p_{ia}p_{ja}) \spc dt   \nonumber\\
\delta (p_{ia}p_{ja}) &=& (p_{ia}' p_{ja}'-p_{ia}p_{ja})
\end{eqnarray}

The 11-dimensional integral in Eq.3 can be reduced to a 3-dimenional integral for 
gaussian distributions, if one notes that in the Rest CS 
$\spc \sigma_{ab}$ depends on
$\vec{v_a}-\vec{v_b}$ and one transforms from the momentum variables
$p_a,p_b$ to two new variables one of which is $\vec{v_a}-\vec{v_b}$. 
This can be done by the transformation
\begin{eqnarray}
\pb_{ia} &=& W_i+\frac{\mu}{m_a} \Delta_i             \nonumber\\
\pb_{ib} &=& W_i-\frac{\mu}{m_b} \Delta_i        \nonumber\\
W_i &=& \frac{p_{ia}+p_{ib}}{\gamma_0 \beta_0 (m_a+m_b)c}   \nonumber\\
\Delta_i &=& \pb_{ia}-\pb_{ib}= \frac{v_{ia}-v_{ib}}{\gamma_0 \beta_0 c}    \nonumber\\
d^3\pb_a d^3\pb_b &=& d^3W d^3\Delta   \nonumber\\
\pb_{ia} &=& \frac{p_{ia}}{\gamma_0 \beta_0 m_ac}   \nonumber\\
\pb_{ib} &=& \frac{p_{ib}}{\gamma_0 \beta_0 m_bc}   \nonumber\\
\frac{1}{\mu} &=& \frac{1}{m_a}+\frac{1}{m_b}                         \nonumber\\
d^3 \pb_a d^3 \pb_b &=& d^3 W d^3 \De            \nonumber\\
\end{eqnarray}

$\Delta_i$ is proportional to the relative velocity, $\vec{v_a}-\vec{v_b} \spc$ 
when the velocities are non-relativistic. A similar transformation 
is used in Ref.1 and Ref.4 except that for them the particles are 
identical and the transformation is simpler.

$\delta (p_{ia}p_{ja})$ can be written as
\begin{eqnarray}
\delta (p_{ia}p_{ja}) &=& p_{ia}q_{ja}+p_{ja}q_{ia}+q_{ia}q_{ja}  \nonumber\\
q_{ia}&=& p_{ia}'-p_{ia}
\end{eqnarray}
This result can written as
\begin{eqnarray}
\delta (\pb_{ia}\pb_{ja}) &=& [(W_{i}\qb_{ja}+W_{j}\qb_{ia}) \muma]+
     [(\muma)^2 (\De_i \qb_{ja}+\De_j \qb_{ia}+\qb_{ia}\qb_{ja})]  \nonumber\\
\qb_{ia} &=& q_{ia} /(\gamma_0 \beta_0 \mu c) \nonumber\\ 
\end{eqnarray}

Eq.3 can be rewritten in terms of $W,\De$ as
\begin{eqnarray} 
<\delta (\pb_{ia}\pb_{ja}) > &=& N_b \int \spc d^3x d^3W
      d^3\De f_a(x,p_a) f_b(x,p_b) |\vec{v_a}-\vec{v_b}|   \nonumber\\
      & &  \sigma_{ab} d\Omega' \spc \delta (\pb_{ia}\pb_{ja}) \spc dt 
                 \nonumber\\
\delta (\pb_{ia}\pb_{ja}) &=& [(W_{i}\qb_{ja}+W_{j}\qb_{ia})\muma]+
     [(\muma)^2 (\De_i \qb_{ja}+\De_j \qb_{ia}+\qb_{ia}\qb_{ja})]  \nonumber\\
\end{eqnarray}
One may note that $\sigma_{ab}$ depends only on $\De$ and not on $W$.
In the expression for $\delta (\pb_{ia}\pb_{ja})$ the second term will be seen 
to depend only on $\De$ and gives rise to the usual intrabeam scattering
growth rate, while the first term depends on $W$ and will be seen to vanish for
identical particles and gives rise to the cooling rates 
for ion electron scattering.

The transformation from $\pv_a,\pv_b$ to $\Wv,\Dev$ allows us to do 
the integral over $d\Omega'$. Eq.7 holds in any CS where the particle
 motion is non-relativistic. For each $\pv_a,\pv_b$ one can define a
center of mass CS, called the CMS, in which $\pv_a+\pv_b=0$.
In the CMS
\[\Delta_i = \pb_{ia}-\pb_{ib}
           =p_{ia}/(\gamma_0 \beta_0 \mu c)   \]
In the CMS, $\Dev$ and $\pv_a$ have the same direction, and $\pv_a$
is scattered by the electrons to $\pv_a \; '$ which is along the direction
given by the polar angles $\theta , \phi$ relative to the direction of
$\pv_a$ or $\Dev$. 

In Eq.7, only the $\qb_{ia}$ depend on the scattering angles $\theta ,\phi \spc$. 
 To do the integral over $d\Omega'$ in the Rest CS one has to evaluate 
the integrals
\begin{eqnarray}
d_i &=& \int d\Omega' \sigma_{ab} \qb_{ia}       \nonumber\\
c_{ij} &=&  \int d\Omega' \sigma_{ab} [(\De_i \qb_{ja}+\De_j \qb_{ia})
         +\qb_{ia}\qb_{ja}]  \nonumber\\
\end{eqnarray}
$d\Omega' \sigma_{ab}$ is an invariant and
$\Dev,\qv_a$ are both the same in the CMS and the Rest CS as they are both the 
difference of 2 vectors that are proportional to a velocity. $d_i$, $c_{ij}$
are tensors in 3-space. If these integrals are  evaluated in the CMS 
and the result is written in terms of tensors in 3-space then the 
result will also hold in the Rest CS.

In the CMS, we introduce a polar coordinate system $\theta,\phi$
where $\theta$ is measured relative to the direction of 
$\vec{p_a}$ or $\vec{\Delta}$ 
and we assume that $\sigma_{ab}(\theta,\phi)$ is a fumction of $\theta$ only.
we can then write
\begin{eqnarray}
\Dev     &=& (0,0,1)|\vec{\Delta}|  \nonumber\\
\vec{p_a} &=& (0,0,1)|\vec{\Delta}| (\gamma_0 \beta_0 \mu c)   \nonumber\\
\vec{p_a \; '} &=& (\sin \theta \cos \phi,\sin \theta \sin \phi,
          \cos \theta)|\vec{\Delta}| (\gamma_0 \beta_0 \mu c)    \nonumber\\
\vec{q_a} &=& (\sin \theta \cos \phi,\sin \theta \sin \phi,
          \cos \theta-1)|\vec{\Delta}|(\gamma_0 \beta_0 \mu c) 
\end{eqnarray}
 In the CMS,
using Eq.9, one finds
\begin{eqnarray}
d_i &=& -2 \pi  \int d\theta sin\theta 
            (1-cos\theta) \sigma_{ab} (0,0,1) |\vec{\Delta}|   \nonumber\\
c_{ij} &=&  \pi \int_{0}^{\pi} d\theta  \sin^3 \theta \sigma_{ab}
     \;|\vec{\Delta}|^2
            \left( \begin{array}{ccr} 1&0&0    \\
                                       0&1&0    \\
                                       0&0&-2    
                                      \end{array} \right )     \nonumber\\
\end{eqnarray}   
In computing $c_{ij}$ one may note that the $\De_i \qb_{ja}+\De_j \qb_{ia}$ 
term in Eq.8 only contributes to $c_{33}$ while the $\qb_{ia}\qb_{ja}$ 
term contributes to to all 3 diagonal elements of $c_{ij}$.
          
These results for $d_i,c_{ij}$ in the CMS can be rewritten in terms of 
tensors in 3-space as
\[  \]
\begin{eqnarray}
d_i &=& -2 \pi  \int d\theta sin\theta  
            (1-cos\theta) \sigma_{ab}   \De_i   \nonumber\\
c_{ij} &=&  \pi \int_{0}^{\pi} d\theta  \sin^3 \theta \sigma_{ab} \;
          (|\vec{\Delta}|^2 \delta_{ij}-3 \De_i \De_j )   \nonumber\\
\end{eqnarray}   
In this form the results will also hold in the Rest CS. Eq. 7 can now 
be rewritten as
\begin{eqnarray} 
<\delta (\pb_{ia}\pb_{ja}) > &=& N_b \int \spc d^3x d^3W
      d^3\De f_a(x,p_a) f_b(x,p_b) |\vec{v_a}-\vec{v_b}|   \nonumber\\
      & & ( [-2\pi \muma (W_i \De_j+W_j \De_i) \int d\theta sin\theta  
            (1-cos\theta) \sigma_{ab} ]_1  \nonumber\\
      & &  +[\pi (\muma)^2  (|\vec{\Delta}|^2 \delta_{ij}-3 \De_i \De_j )  
           \int d\theta  sin^3 \theta \sigma_{ab}]_2 ) \spc dt \nonumber\\       
\end{eqnarray}

Eq.12 can be used to compute either  intrabeam scattering 
for identical particles or 
electron cooling. If the $a$ and $b$ particles are identical, then the second term
indicated by $[\spc ]_2$ and called the $\De$-term gives the growth rates for 
intrabeam scattering. In this case, the first term, indicated by $[\spc ]_1$
and called the W-term, will vanish. This is shown below for gaussian distributions
and also can be shown to hold for any distribution because of the symmetry 
of the $a$ and $b$ particles. If the $b$ particle is much 
lighter than the $a$ particle,
the W-term gives rise to cooling of the $a$ particles and the $\De$-term 
is smaller than the W-term by the factor $m_b/m_a$. This is shown below 
for gaussian distributions. Eq. 12 holds for any distibutions, 
$f_a(x,p_a), f_b(x,p_b)$. In the next section, we will specialize to gaussian
distributions.

it is often assumed that $\sigma_{ab}$ is given by the Coulonb cross-section in 
the CMS CS for the $a$ and $b$ particles. This is given by
\begin{eqnarray}
\sigma_{ab} &=& (\frac {r_{ab}} {\beta_{ab}^2})^2 \frac{1}{(1-cos \theta)^2} 
                    \nonumber\\
r_{ab} &=& \frac{Z_aZ_b e^2}{\mu c^2}     \nonumber\\
\beta_{ab} c &=& |\vec{v_a}-\vec{v_b}|   \nonumber\\
\end{eqnarray} 

The integrals over $\theta$ in Eq.12 can then be written as
\[     \]
\begin{eqnarray}
\int d\theta sin\theta (1-cos\theta) \frac{1}{(1-cos \theta)^2} &=& 
      ln \lbr 1+\left (\frac{\beta_{ab}^2 b_{maxab}}{r_{ab}} \right )^2\rbr   \nonumber\\
\int d\theta  sin^3 \theta \frac{1}{(1-cos \theta)^2} &=& 2\lbr  
      ln \lbr 1+\lpar \frac{\beta_{ab}^2 b_{maxab}}{r_{ab}}\rpar^2 \rbr 
      -\frac{1}{1+( r_{ab}/(\beta_{ab}^2 b_{maxab}))^2} \rbr         \nonumber\\
tan(\theta_{min}) &=& \frac{r_{ab}}{\beta_{ab}^2 b_{maxab}} \nonumber\\
\end{eqnarray} 
$b_{maxab}$ is the maximun allowed impact parameter in the CMS. $\theta_{min}$
is the smallest allowed scattering angle in the CMS.

It will be seen below that to compute the cooling rates for the emittances
one will also need the cooling rates for $<x_{ia}p_{ja}>$. When the $a$ and $b$
particles are identical, the  $<x_{ia}p_{ja}>$ are zero , but not zero 
when the particles are different. Using Eq.7, one finds 
\begin{eqnarray} 
<\delta (x_{i}\pb_{ja}) > &=& N_b \int \spc d^3x d^3W
      d^3\De f_a(x,p_a) f_b(x,p_b) |\vec{v_a}-\vec{v_b}|   \nonumber\\
      & &  \sigma_{ab} d\Omega' \spc \delta (x_{ia}\pb_{ja}) \spc dt 
                 \nonumber\\
\delta (x_{i}\pb_{ja}) &=& x_{i} \delta \pb_{ja}= x_{i} \qb_{ja} \muma
        \nonumber\\
\mbox{From Eq.11 one has}   & &               \nonumber\\
\int d\Omega' \sigma_{ab} \qb_{ja} &=& -2 \pi  \int d\theta sin\theta  
            (1-cos\theta) \sigma_{ab}   \De_j   \nonumber\\
\mbox{which gives}   & &               \nonumber\\
<\delta (x_{i}\pb_{ja}) > &=& N_b \int  d^3x d^3W
      d^3\De \spc f_a(x,p_a) f_b(x,p_b) |\vec{v_a}-\vec{v_b}|   \nonumber\\
      & &   [-2 \pi  \int d\theta sin\theta  
            (1-cos\theta) \sigma_{ab} \spc x_{i}   \De_j \muma]
          \spc dt 
                 \nonumber\\
\end{eqnarray}

Eq.15 shows that $<\delta (x_{i}\pb_{ja}) >$ gives rise to a cooling term 
which vanishes when the particles are identical , or when $\alpha_i=0$ for the
ion partcle for a gaussian distribution.

\section{Cooling  rates for $<p_{ia}p_{ja}>$ in the Rest CS for Gaussian 
     distributions}

In this section, we will find the cooling rates due to the scattering 
of the ions by the electrons in the cooling section when the ion and 
electron bunches have gaussian distributions. In Eq.12 , we will keep only the 
$W$-term as the $\Delta-$term , discussed later, is smaller by the factor
$m_b/m_a$ In this paper, it will be assumed that the dispersion is zero in
the cooling section.

For a gaussian distribution, $f_a(x,p_a)$ is given  for the ion bunch for zero dispersion by Ref.[3],
\begin{eqnarray} 
f_a(x,p_a) &=& \frac{1}{\Gamma_a} exp[-S_a(x,p_a)]  \nonumber\\
\Gamma_a &=& \int d^3xd^3p \;  exp[-S_a (x,p_a)]   \nonumber\\
\Gamma_a &=& \pi^3 \epb_{xa} \epb_{sa} \epb_{ya}    \nonumber\\
\end{eqnarray}
%
\begin{eqnarray}
S_a &=& S_{xa}+S_{ya}+S_{sa}      \nonumber\\ 
  & &                  \nonumber\\ 
S_{xa} &=& \frac{1}{\bar{\epsilon_{xa}}}  \epsilon_{xa} (x,x'_a) \spc 
        x_a'=p_{xa}/p_{0a}             \nonumber\\
\epsilon_{xa} (x,x'_a) &=& [x^2+(\beta_x x'_a+\alpha_{xa} x)^2]/\beta_{xa} 
                       \nonumber\\
  & &                  \nonumber\\ 
S_{ya} &=& \frac{1}{\bar{\epsilon_{ya}}}  \epsilon_{ya} (y,y'_a) \spc 
        y_a'=p_{ya}/p_{0a}             \nonumber\\
\epsilon_{ya} (y,y'_a) &=& [y^2+(\beta_y y'_a+\alpha_{ya} y)^2]/\beta_{ya} 
                       \nonumber\\
  & &                  \nonumber\\
S_s &=& \frac{1}{\bar{\epsilon_s}}  \epsilon_s (s,p_s/p_{0a})  \nonumber\\
\epsilon_s (s,p_s/p_{0a})&=& \frac{s^2}{2\sigma_s^2}+\frac{(p_s/p_{0a})^2}
        {2 \sigma_p^2} \nonumber\\
\epsilon_s (s,p_s/p_{0a}) &=& \frac{1}{\beta_s} (s)^2+\beta_s (p_s/p_{0a})^2 
                \nonumber\\
\epsilon_s (s,p_s/p_{0a}) &=& [(s)^2+(\beta_s (p_s/p_{0a}))^2]/\beta_s \nonumber\\
\beta_s &=& \sigma_s/\sigma_p  \nonumber\\
\bar \epsilon_s &=& 2 \sigma_s \sigma_p  
\end{eqnarray}
%
A longitudinal emittance has been introduced
so that the longitudinal motion and the transverse motions can be treated in 
a similar manner. $\beta_s$ in the Rest CS is larger than $\beta_s$ 
in the Laboratory CS by the factor $\gamma_0^2$. $s,p_s$ are the paricle 
longitudinal position and momentum in the Rest CS.

In Eq.12 we will now do the integration over $d^3x d^3W$ using the above gaussian 
ditributions. Because there is no dispersion in the cooling section the 
integral over $dxdW_x$ or $dsdW_s$ or $dydW_y$ can each be treated 
in a similar way.
Eq.12 can now be written using the Coulomb cross-section as
\begin{eqnarray} 
\delta <(\pb_{ia}\pb_{ja}) > &=& \frac {N_b}{\Gamma_a \Gamma_b}\int \spc d^3x d^3W
      d^3\De exp[-(S_a+S_b)] |\vec{v_a}-\vec{v_b}|   \nonumber\\
      & & \muma \Wb_{ij} \spc
      (\frac {r_{ab}} {\beta_{ab}^2})^2 \spc
      ln \lbr 1+\lpar \frac{\beta_{ab}^2 b_{maxab}}{r_{ab}}\rpar^2 \rbr 
      \spc dt \nonumber\\
\Wb_{ij} &=& -2 \pi (W_i \De_j+W_j \De_i)
\end{eqnarray}
We rewrite $S_a+S_b$ as
\begin{eqnarray}
         S_a+S_b &=&  \Sigma_i(S_{ia}+S_{ib}) \spc i=x,y,s  \nonumber\\ 
S_{ia} &=& \frac {1} {\epb_{ia}} 
   \lbr  
    \frac{x_{ia}^2}{\beta_{ia}}+
   ( \be_{ia}^{1/2} \pb_{ia}+\frac {\al_{ia} x_{ia}} {\be_{ia}^{1/2}} )^2 
   \rbr   
   \nonumber\\
S_{ia} &=& \frac {1} {\epb_{ia}} 
   \lbr  
    \frac{x_{ia}^2}{\beta_{ia}}+
   ( \be_{ia}^{1/2} (W_i+\frac{\mu}{m_a} \Delta_i) +\frac {\al_{ia} x_{ia}} 
     {\be_{ia}^{1/2}} )^2 
   \rbr   
   \nonumber\\
                    & &                            \nonumber\\
S_{ia}+S_{ib} &=& A_{11i} x_i^2 +A_{22i} W_i^2 +2 A_{12i} x_i W_i 
           +(A_{10i} x_i +A_{01i} W_i) \De_i +A_{00i} \De_i^2          \nonumber\\
                    & &                            \nonumber\\
A_{11i} &=& \lbr \frac{1+\al_{i}^2}{\be_{i} \epb_{i}} \rbr_+  \spc   
         A_{22i}=\lbr \frac{\be_{i}}{\epb_{i}} \rbr_+               \nonumber\\
A_{12i} &=& \lbr \frac{\al_{i}}{\epb_{i}} \rbr_+     \spc
         A_{10i}=\lbr 2 \mum \frac{\al_{i}}{\epb_{i}} \rbr_-      \nonumber\\
A_{01i} &=& \lbr 2 \mum \frac{\be_{i}}{\epb_{i}} \rbr_-      \spc
        A_{00i}=\lbr (\mum)^2 \frac{\be_{i}}{\epb_{i}} \rbr_+  \nonumber\\
\end{eqnarray}

The symbols $[ (\spc ) ]_+ $ and $[ (\spc ) ]_-$ are defined by
\[[ (\spc ) ]_+ =(\spc)_a+(\spc)_b \]
\[[ ( \spc) ]_- =(\spc)_a-(\spc)_b \]
We will now make a transformation to eliminate the $2 A_{12i} x_i W_i$ term
in $S_{ia}+S_{ib}$. We rewrite $S_{ia}+S_{ib}$ as
\begin{eqnarray}
S_{ia}+S_{ib}  &=& A_{11i} x_i^2 +A_{22i} W_i^2 +2 A_{12i} x_i W_i 
                  +(A_{10i}x_i +A_{01i} W_i) \De_i +A_{00i} \De_i^2   \nonumber\\
               &=& [A_{11} x^2 +A_{22} W^2 +2 A_{12} x W 
                  +(A_{10}x +A_{01} W) \De +A_{00} \De^2]_i   \nonumber\\
               &=& [x^2(A_{11}-\frac{A_{12}^2}{A_{22}})+
                   (A_{22}^{1/2}W+\frac{A_{12}}{A_{22}^{1/2}} x)^2 \nonumber\\
              & &    +(A_{10}x +A_{01} W) \De +A_{00} \De^2   ]_i \nonumber\\
              & &                    \nonumber\\
\eta_i &=& \lbr \frac {\Ab^{1/2}}{A_{22}^{1/2}} x \rbr_i \spc \spc
         p_{\eta i} = \lbr A_{22}^{1/2}W+\frac{A_{12}}{A_{22}^{1/2}} x\rbr_i \nonumber\\
\Ab_i &=& [A_{11}A_{22}- A_{12}^2]_i   \nonumber\\
x_i &=& [x_{\eta}\eta]_i \spc \spc W_i=[(W_{\eta}\eta+W_{p_{\eta}} p_{\eta} )]_i 
      \nonumber\\
dx_i dw_i &=& \lbr \frac{1}{\Ab^{1/2}} d\eta dp_{\eta}\rbr_i \nonumber\\
x_{\eta i} &=& \lbr \frac{A_{22}^{1/2}}{\Ab^{1/2}}\rbr_i \spc \spc W_{\eta i}=    
  \lbr  -\frac {A_{12}}{\Ab^{1/2}} \rbr_i 
\spc \spc  W_{p_{\eta i}}= \lbr \frac{1}{A_{22}^{1/2}}\rbr_i 
\nonumber\\
              & &                    \nonumber\\
S_{ia}+S_{ib}  &=& [\eta^2+p_{\eta}^2
                  +(A_{10}x +A_{01} W) \De +A_{00} \De^2   ]_i \nonumber\\
              &=& [\eta^2+p_{\eta}^2
                  +(B_{10}\eta +B_{01}p_{\eta} ) \De +A_{00} \De^2 ]_i \nonumber\\
B_{10i} &=& [A_{10}x_{\eta}+A_{01} W_{\eta}]_i  
          \spc \spc B_{01i}= [A_{01} W_{p_{\eta}}]_i  
        \nonumber\\
B_{10i} &=& \lbr A_{10} \frac{A_{22}^{1/2}}{\Ab^{1/2}}
    -A_{01} \frac {A_{12}}{\Ab^{1/2}}\rbr_i  
          \spc \spc B_{01i}= \lbr A_{01} \frac{1}{A_{22}^{1/2}}\rbr_i  
        \nonumber\\
              & &                    \nonumber\\
\Wb_{ij} &=& -2 \pi [ 
           (W_{\eta}\eta +W_{p_{\eta}} p_{\eta})_i \De_j
          +(W_{\eta}\eta +W_{p_{\eta}} p_{\eta})_j\De_i 
                    ]  \nonumber\\
\end{eqnarray}

In the expression for $S_{ia}+S_{ib}$ given  at the end of Eq.19, 
the linear terms in $\eta,p_{\eta}$ can be eliminated by the transformation
\begin{eqnarray}
\etb_i &=& \lbr \eta+\frac{B_{10}}{2} \De\rbr _i \spc \spc 
\petb_i=\lbr p_{\eta}+\frac{B_{01}}{2} \De\rbr _i      \nonumber\\
S_{ia}+S_{ib}  &=& [\etb^2+\petb^2
                   +(A_{00}-B_{10}^2/4-B_{01}^2/4) \De^2 ]_i \nonumber\\
\Wb_{ij} &=& -2 \pi [  
           [W_{\eta}(\etb-\frac{B_{10}}{2}\De)]_i \De_j +[W_{p_{\eta}} 
             (\petb-\frac{B_{01}}{2} \De)]_i \De_j    \nonumber\\ 
     & & +[W_{\eta}(\etb-\frac{B_{10}}{2}\De)]_j \De_i +[W_{p_{\eta}} 
             (\petb-\frac{B_{01}}{2} \De)]_j \De_i
                    ]   \nonumber\\
\end{eqnarray}

Eq.17 can now be rewritten as
\begin{eqnarray} 
<\delta (\pb_{ia}\pb_{ja}) > &=& \frac {N_b}{\Gamma_a \Gamma_b} 
       \frac{1}{\Ab^{1/2}_p}   
       \int \spc d^3\etb d^3\petb d^3\De \spc exp[-(S_a+S_b)] 
        |\vec{v_a}-\vec{v_b}|  
       \nonumber\\
      & & \muma \Wb_{ij} (\frac {r_{ab}} {\beta_{ab}^2})^2 \spc 
       ln \lbr 1+\lpar \frac{\beta_{ab}^2 b_{maxab}}{r_{ab}}\rpar^2 \rbr  
      \spc dt 
     \nonumber\\
\Wb_{ij} &=& -2 \pi [ 
           [W_{\eta}(\etb-\frac{B_{10}}{2}\De)]_i \De_j +[W_{p_{\eta}} 
             (\petb-\frac{B_{01}}{2} \De)]_i \De_j    \nonumber\\ 
     & & +[W_{\eta}(\etb-\frac{B_{10}}{2}\De)]_j \De_i +[W_{p_{\eta}} 
             (\petb-\frac{B_{01}}{2} \De)]_j \De_i
                    ]  \nonumber\\
\Ab^{1/2}_p &=& \Ab_x^{1/2}\Ab_y^{1/2}\Ab_s^{1/2}   \nonumber\\
\end{eqnarray}
Using Eq.20 for $S_{ia}+S_{ib}$ and for $\Wb_{ij}$ , one can do the 
integral over $d^3\etb d^3\petb$ and get
\[  \]
\[  \]
\[  \]
\[  \]
\begin{eqnarray} 
\delta <(\pb_{ia}\pb_{ja}) > &=& \frac {N_b}{\Gamma_a \Gamma_b} 
       \frac{1}{\Ab^{1/2}_p} \pi^3 r_{ab}^2 c  \muma \Wh_{ij} \nonumber\\
      & &  
       \int \spc  d^3\De \frac{ exp[-(\la_x \De_x^2+\la_y \De_y^2+\la_s \De_s^2)]}
       {\beta_{ab}^3}   \De_i\De_j
       \nonumber\\
      & &  
       ln \lbr 1+\lpar \frac{\beta_{ab}^2 b_{maxab}}{r_{ab}}\rpar^2 \rbr  
      \spc dt 
     \nonumber\\
              & &                    \nonumber\\
\Wh_{ij} &=& 2 \pi \lbr 
           (W_{\eta}\frac{B_{10}}{2} +W_{p_{\eta}} \frac{B_{01}}{2})_i
          +(W_{\eta}\frac{B_{10}}{2} +W_{p_{\eta}} \frac{B_{01}}{2})_j 
                  \rbr  \nonumber\\
      & &  \nonumber\\
\beta_{ab} &=& \gamma_0 \beta_0 (\De_x^2+\De_y^2+\De_s^2)^{1/2} \nonumber\\ 
\lambda_i &=& \lbr A_{00}-(\frac{B_{10}}{2})^2-(\frac{B_{01}}{2})^2\rbr_i \nonumber\\
\Ab_i &=& [A_{11}A_{22}- A_{12}^2]_i   \nonumber\\
\Ab^{1/2}_p &=& \Ab_x^{1/2}\Ab_y^{1/2}\Ab_s^{1/2}   \nonumber\\
              & &                    \nonumber\\
x_{\eta i} &=& \lbr \frac{A_{22}^{1/2}}{\Ab^{1/2}}\rbr_i \spc \spc W_{\eta i}=    
  \lbr  -\frac {A_{12}}{\Ab^{1/2}} \rbr_i 
\spc \spc  W_{p_{\eta i}}= \lbr \frac{1}{A_{22}^{1/2}}\rbr_i 
\nonumber\\
B_{10i} &=& [A_{10}x_{\eta}+A_{01} W_{\eta}]_i  
          \spc \spc B_{01i}= [A_{01} W_{p_{\eta}}]_i  
        \nonumber\\
B_{10i} &=& \lbr A_{10} \frac{A_{22}^{1/2}}{\Ab^{1/2}}
    -A_{01} \frac {A_{12}}{\Ab^{1/2}}\rbr_i  
          \spc \spc B_{01i}= \lbr A_{01} \frac{1}{A_{22}^{1/2}}\rbr_i  
        \nonumber\\
              & &                    \nonumber\\
A_{11i} &=& \lbr \frac{1+\al_{i}^2}{\be_{i} \epb_{i}} \rbr_+  \spc   
         A_{22i}=\lbr \frac{\be_{i}}{\epb_{i}} \rbr_+               \nonumber\\
A_{12i} &=& \lbr \frac{\al_{i}}{\epb_{i}} \rbr_+     \spc
         A_{10i}=\lbr 2 \mum \frac{\al_{i}}{\epb_{i}} \rbr_-      \nonumber\\
A_{01i} &=& \lbr 2 \mum \frac{\be_{i}}{\epb_{i}} \rbr_-      \spc
        A_{00i}=\lbr (\mum)^2 \frac{\be_{i}}{\epb_{i}} \rbr_+  \nonumber\\
\end{eqnarray}

Eq.23 is our final result for the cooling rates for $<p_{ia}p_{ja}>$ 
in the Rest CS, for two overlapping 
gaussian bunches , with no dispersion in the cooling section.
For this case one gets zero results when $i \neq j$. The remaining 
3-dimensional integral over $d^3\De$ is an integral over the relative
velocities of the ions and electrons.

It will be seen below that to compute the cooling rates for the emittances
one will also need the cooling rates for $<x_{ia}p_{ja}>$. For gaussian 
distributioins, using the coulomb cross section and Eq.15, Eq.18 is replaced by
\begin{eqnarray} 
\delta <(x_i\pb_{ja}) > &=& \frac {N_b}{\Gamma_a \Gamma_b}\int \spc d^3x d^3W
      d^3\De exp[-(S_a+S_b)] |\vec{v_a}-\vec{v_b}|   \nonumber\\
      & & \muma \xb_{ij} \spc
      \lbr \frac {r_{ab}} {\beta_{ab}^2}\rbr^2 \spc
      ln \lbr 1+\lpar \frac{\beta_{ab}^2 b_{maxab}}{r_{ab}}\rpar^2 \rbr  
      \spc dt \nonumber\\
\xb_{ij} &=& -2 \pi x_i \De_j
\end{eqnarray}

After going from the $x,W$ coordinates to $\eta,p_{\eta}$ and integrating over 
$\eta,p_{\eta}$ Eq.23 is replaced by 
\begin{eqnarray} 
\delta <(x_i\pb_{ja}) > &=& \frac {N_b}{\Gamma_a \Gamma_b} 
       \frac{1}{\Ab^{1/2}_p} \pi^3 r_{ab}^2 c  \muma \xh_{ij} \nonumber\\
      & &  
       \int \spc  d^3\De \frac{ exp[-(\la_x \De_x^2+\la_y \De_y^2+\la_s \De_s^2)]}
       {\beta_{ab}^3}   \De_i\De_j
       \nonumber\\
      & &  
       ln \lbr 1+\lpar \frac{\beta_{ab}^2 b_{maxab}}{r_{ab}}\rpar^2 \rbr  
      \spc dt 
     \nonumber\\
              & &                    \nonumber\\
\xh_{ij} &=& 2 \pi \lbr x_{\eta}\frac{B_{10}}{2}\rbr_i 
                                  \nonumber\\
              & &                    \nonumber\\
x_{\eta i} &=& \lbr \frac{A_{22}^{1/2}}{\Ab^{1/2}}\rbr_i \spc \spc W_{\eta i}=    
  \lbr  -\frac {A_{12}}{\Ab^{1/2}} \rbr_i 
\spc \spc  W_{p_{\eta i}}= \lbr \frac{1}{A_{22}^{1/2}}\rbr_i 
\nonumber\\
B_{10i} &=& [A_{10}x_{\eta}+A_{01} W_{\eta}]_i  
          \spc \spc B_{01i}= [A_{01} W_{p_{\eta}}]_i  
        \nonumber\\
B_{10i} &=& \lbr A_{10} \frac{A_{22}^{1/2}}{\Ab^{1/2}}
    -A_{01} \frac {A_{12}}{\Ab^{1/2}}\rbr_i  
          \spc \spc B_{01i}= \lbr A_{01} \frac{1}{A_{22}^{1/2}}\rbr_i  
        \nonumber\\
\end{eqnarray}

\section{Emittance growth rates}

One can compute growth rates for the average emittances, $<\epsilon_{ia}>$ in the 
Laboratory Coordinate System, from the growth rates for $<p_{ia}p_{ja}>$ in the 
Rest Coordinate System. In the following , $dt$ is the time interval in the
Laboratory System and $d\tilde{t}$ is the time interval in the Rest System.
$dt=\gamma d\tilde{t}$. The final results are, for zero dispersion,
\begin{eqnarray}
\frac{d}{dt} <\epb_{ia}> &=& \frac{\beta_{ia}}{\gamma} \frac{d}{d\tilde{t}}
        <\pb_{ia}^2>  
         +  \frac{2 \alpha_{ia} }{\gamma} \frac{d}{d\tilde{t}} <x_i \pb_{ia}> 
                   \spc i=x,y,s     \nonumber\\
\end{eqnarray}

To derive the above results, the simplest case to treat is that of the vertical 
emittance. The verical emmitance is given by
\begin{eqnarray}
\epb_{ya}(y,y_a') &=& [y^2+(\beta_{ya} y'_a+\alpha_{ya} y)^2]/\beta_{ya} 
    \spc y_a'=\pb_{ya}          \nonumber\\
\delta \epb_{ya} &=& \beta_{ya}  \delta (\pb_{ya}^2)
        +\delta (2 \alpha_{ya} y  (\pb_{ya})         \nonumber\\
\frac{d}{dt} <\epb_{ya}> &=& \frac{\beta_{ya}}{\gamma} \frac{d}{d\tilde{t}}
        <\pb_{ya}^2>  
         +  \frac{2 \alpha_{ya} }{\gamma} \frac{d}{d\tilde{t}} <y \pb_{ya}> 
                        \nonumber\\
\end{eqnarray}
In Eq.(27), $y_a'=\pb_{ya}$, $\delta \epsilon_{ya}$ is the change in 
$\epb_{ya}$ in a scattering event. Similar results will hold for 
$\epb_{xa}$ and $\epb_{sa}$ for zero dispersion.

\section*{The $\Delta$ term in electron cooling}

In the previous section it was assumed that in Eq.12 one could drop the 
second term or $\De$ term compared to the first term or $W$ term.This is true when
$m_b<<m_a$ and $\pb_a \simeq \pb_b$ in the Rest CS. Using Eq.4, one can write
\begin{eqnarray}
W_i &=& \lbr \pb_a \frac{m_a}{m_a+m_b}+\pb_b \frac{m_b}{m_a+m_b} \rbr_i \nonumber\\
\De_i &=& [\pb_a-\pb_b]_i                        \nonumber\\
      & &                    \nonumber\\
W_i &\simeq & [ \pb_a]_i           \nonumber\\
\De_i &=& [\pb_a-\pb_b]_i                        \nonumber\\
\end{eqnarray}
Thus $W$ and $\De$ are both of the same order as $\pb_a$ . If the motion is 
non-relativistic in the Rest CS, $\qb_a \simeq \De \simeq \pb_a $. From this it
follows that the $\De$ term in Eq.12 is smaller than the $W$ term by the factor 
$m_b/m_a$.

It has also been assumed that the motion in the Rest CS is non-relativistic. 
In the Laboratory CS, the rms spread in the relative momentum is given by 
\begin{eqnarray}
\sigma_{pi} &=& \lbr \frac{\epb_i}{2 \beta_i}\rbr^{1/2} \spc i=x,y,s \nonumber\\
\end{eqnarray}
For gold ions in RHIC  at $\gamma=100$
\[\spc \epb_x=\epb_y=5e-8,\beta_x=\beta_y=50 \spc and \spc \sigma_{px}=\sigma_{py}=2.24e-5  \]
\[\spc \epb_s=1.8e-4,\beta_s=300m \spc and \spc \sigma_{ps}=.55e-3  \]
In the Rest CS $\sigma_{px},\sigma_{py}$ are unchanged at 2.24e-5   And 
$\sigma_{ps}$ is reduced by the factor $\gamma$ to .55e-5  The spread in 
each of the momenta in the Rest CS is of the order of $1e-3 m_ac$ 
since $\gamma=100$ and the 
ion velocities are of the order of 1e-3c. Similar numbers hold for the 
electrons in the electron bunch.

\section*{References}

\noindent
1. A. Piwinski Proc. 9th Int. Conf. on High Energy Accelerators (1974) 405

\noindent
2. G. Parzen BNL report C-A/AP/N0.150 (2004) 

     and at http://arxiv.org/ps\uu cache/physics/pdf/0405/0405019.pdf

\noindent
3. G. Parzen BNL report C-A/AP/N0.169 (2004)

   and at http://arxiv.org/ps\uu cache/physics/pdf/0410/0410028.pdf

\noindent
4. J.D. Bjorken and S.K. Mtingwa, Part. Accel.13 (1983) 115

\end {document}